\documentstyle[12pt]{article}

\newcounter{tabl}
\newcommand{\be}{\begin{equation}}
\newcommand{\ee}{\end{equation}}
\newcommand{\beq}{\begin{eqnarray}}
\newcommand{\eeq}{\end{eqnarray}}
\newcommand{\bea}[2]{\be\label{#2}\begin{array}{#1}}
\newcommand{\eea}{\end{array}\ee}

\def\Nb{{\rm \bf N}}
\def\Rb{{\rm \bf R}}

\def\({\left(}
\def\){\right)}
\def\[{\left[}
\def\]{\right]}

\def\11{1\!\! 1}

   \def\CA {{\cal A}}

\newcommand{\Ah}{A_{\rm hor}}

\newcommand{\Ref}[1]{(\ref{#1})}

\begin{document}
%
%

\title{
On the counting of black hole states \\
in loop quantum gravity}

\author{Sergei Alexandrov\thanks{email: S.Alexandrov@phys.uu.nl}
}

\date{}

\maketitle

\vspace{-1cm}

\begin{center}
\it  Institute for Theoretical Physics \& Spinoza Institute, \\
Utrecht University, Postbus 80.195, 3508 TD Utrecht, The Netherlands
\end{center}

\vspace{0.1cm}

\begin{abstract}
We argue that counting black hole states in loop quantum gravity
one should take into account only states with the minimal spin at the horizon.
\end{abstract}

\vspace{1.2cm}

One of the main achievements of the loop approach \cite{loops1,loops2,Rov,Rov-dif} 
to quantization of gravity
is the derivation of the Bekenstein--Hawking entropy of a spherically symmetric 
black hole via counting of micro states. The original idea was proposed in \cite{bhentr1}
and the modern version of this calculation
was developed in \cite{ABCK} (see also \cite{ACK,ABK} for a detailed derivation). 
Using the framework of the so called isolated horizons, 
the problem was reduced to the problem of counting of the number 
of different ways to puncture a two-dimensional sphere,
representing the horizon, which give the horizon area close to a given value.

More precisely, each puncture is equipped with a spin $j$ labeling irreducible
representations of SU(2) which is considered to be the gauge group of the theory.
The puncture contributes the amount given by 
the eigenvalue of the area operator \cite{area,ALarea} 
\be
A_j= 8\pi G\hbar\, \beta \sqrt{C(su(2))}
= 8\pi G \hbar\, \beta \sqrt{ j(j+1) }
\label{assu}
\ee 
to the total area of the horizon.
Here $\beta$ is the so called Immirzi parameter \cite{imir} which is fixed
by the requirement that one obtains the Bekenstein--Hawking formula 
with the correct coefficient 1/4.
The crucial point was that each puncturing of the horizon has a weight 
related to the dimension of an internal space.
This space is the space of all flat U(1) connections on the punctured sphere 
whose holonomies around the punctures are given by $e^{4\pi i m/k}$, where 
the parameter $k={\Ah\over 4\pi G\hbar \, \beta}$ 
plays the role of the ``level" of the U(1) Chern--Simons theory
living on the horizon and $m$ is a half-integer appearing 
in the decomposition of the representation $j$ of SU(2) associated
with a puncture onto U(1).
In the limit of a large number of punctures 
one can think simply that each puncture is equipped with an internal space
of the following dimension
\be
\dim H_j=2j+1 .
\label{dim}
\ee

In \cite{ABCK} it was argued that the leading contribution to the entropy, when
the horizon area is large and the black hole can be considered quasiclassically,
comes from the puncturing by only spins 1/2 and thus the number of micro states
is approximately the dimension of the internal space associated with this puncturing.
On the other hand, recently it was noticed \cite{Lew,Meis} 
that in the original paper \cite{ABCK} there was a mistake 
and the contributions from all spins can not be neglected.
In any case, we see that in the original formulation of the problem
one takes into account all possible sets of punctures giving the same area. 

In this note we argue that this idea is not correct. 
In fact, from the very beginning one should restrict oneself only to the 
punctures associated with the minimal spin, or more generally, 
with the representation minimizing the area operator.

Our argument is the following.  Let us consider two different ways to puncture
the horizon which give the same area. In the original approach \cite{ABCK}
they were considered as two {\it micro} states of the same black hole.
But in fact if the two sets of punctures are different, the bulk geometries
(spacetime outside the black hole), which can be consistently glued
to the horizon, are also different, at least in its neighborhood. 
Indeed, the bulk geometry is encoded into the spin network representing 
a quantum state. The difference in punctures ensures that
the two spin networks are also different and hence they describe
different spacetimes.

Due to this, we should rather consider such states as describing two different 
black holes than two micro states of the same black hole. 
In other words, counting different micro states of a black hole, 
one should ensure that we describe the same state outside its horizon
and, when one punctures the horizon in two different ways, 
one does not satisfy this requirement. 
Thus, one should fix a set of punctures from the very beginning.

One can further develop this picture what also allows 
to understand which punctures of the horizon should be considered in the
calculation of the entropy.
Let us think of the different sets of punctures as different
{\it macro} states of the black hole. With each macro state one can associate
an entropy. Accordingly to our reasoning, the entropy coincides with the logarithm
of the dimension of the internal space assigned to the punctured horizon. 
Approximately, it is given by   
\be
S(\{j_p\})=\log N(\{j_p\})
\approx\log\(\prod\limits_{{\rm punctures}\ p}\dim H_{j_p}\),
\label{entlog}
\ee
where the spins $j_p$ are subject to the condition
\be
\sum\limits_{{\rm punctures}\ p}A_{j_p}=\Ah
\label{aj}
\ee
with $A_j$ being the area \Ref{assu} associated with each puncture.
Since the system always evolves to the state with the maximal entropy, one should consider
only that set of punctures which gives the maximum to the expression \Ref{entlog}.
It is clear that it is realized by the set $\{j_p=1/2\}$ where the number of punctures 
is $n={\Ah / A_{1/2}}$.
Thus, the final state of the evolution, or the equilibrium state, 
corresponds to the horizon with all punctures equipped with spin $1/2$, whereas 
the puncturing by higher spins can be viewed as an analogue of
an excited state of the black hole.   
Due to this the entropy is given entirely by the dimension of the total internal space
associated with the punctures by the minimal spin
and reproduces the familiar result of \cite{ABCK}
\be
S=\log N(1/2)={\beta_0\over\beta}{\Ah\over 4G\hbar}, 
\qquad \beta_0={\log 2 \over \pi \sqrt{3} }.
\label{entsu}
\ee

This statistical picture is confirmed also by the following observation.
Since the bulk geometries corresponding to the two different sets of punctures are different
as well, the local geometry of the horizon itself
also differs in the two cases. It can be arbitrarily complicated depending
on the distribution of spins over the punctures.  
As a result, in the continuum limit the horizon in general is not spherically
symmetric and it has some complicated geometry. 

On the other hand, it is known that approaching the equilibrium the black hole looses
all its ``hairs" and, if there is no angular momentum, it becomes spherically symmetric.
Thus, all non-homogeneity of the horizon geometry should disappear. In terms of loop quantum 
gravity this means that the interaction changes the punctures on the horizon
in such a way that their distribution becomes homogeneous, given by a single spin value. 
Correspondingly, the horizons with a non-homogeneous distribution of spins 
over the punctures describe non-equilibrium black holes without spherical symmetry.
Finally, it is natural to think that $n$ punctures 
by spin $1/2$ give a better approximation for 
the spherically symmetric continuum limit than $\sim {n/j}$ punctures by spin $j$.
This allows to choose the value of the spin common for all punctures and relevant 
for the calculation of the entropy to be $j=1/2$.

\bigskip\bigskip

The previous reasoning does not actually depend on the gauge group which is used to label
the edges of spin networks and punctures of the horizon. It is quite general and 
can be always applied as soon as there is a non-vanishing ``quantum" of area
given by the minimal eigenvalue of the area operator.
In particular, one can try to extend the previous derivation of the black hole entropy
to the covariant formulation of loop quantum gravity developed in 
\cite{SA,AV,SAcon,SAhil,AlLiv}. 
This formulation is based on the Lorentz gauge group SO(3,1) and
predicts a different spectrum for the area operator \cite{AV,SAcon}
\beq
{\CA}&=& 8\pi G\hbar \sqrt{C(so(3)) -C_1(so(3,1))}
\nonumber \\
&=& 8\pi G\hbar \sqrt{ j(j+1) +\rho^2 - n^2+1}, 
\label{as}
\eeq
where $j\ge n$, $2n\in \Nb$ and $\rho\in \Rb$, and probably some additional conditions
should be imposed on the admissible values of the labels \cite{SAhil,AlLiv}. 
In contrast to \Ref{assu}, 
this spectrum is continuous, but what is important that 
it leads to a non-vanishing area quantum. It is obtained choosing $j=n=\rho=0$
so that $\CA_{\rm min}=8\pi G\hbar$.
Then the remaining problem to reproduce the black hole entropy is 
to understand what internal space should be associated to the horizon 
with $n$ punctures, labeled by the $n=\rho=0$ representation of SO(3,1),\
and to find the asymptotics of its dimension in the limit of a
large number of punctures, where
\be
n={\Ah\over \CA_{\rm min}}={\Ah\over 8\pi G \hbar}.
\label{nnn}
\ee 
To get the correct Bekenstein--Hawking formula one needs the following asymptotics
\be
\dim H\sim e^{2\pi n}.
\label{asym}
\ee
Whether one can introduce an internal space with the property \Ref{asym} 
will be investigated elsewhere.

\section*{Acknowledgements}

The author is grateful to Kirill Krasnov and especially to Carlo Rovelli
for very valuable discussions. Also it is a pleasure to thank the organizers
of the Workshop ``Loops and Spinfoams" in Marseille, where
this work was initiated, for the kind hospitality.

\end{document}